\documentclass[12pt]{article}
\usepackage{graphicx}
\usepackage[bookmarksnumbered,colorlinks,plainpages]{hyperref}

\begin{document}

\title{About ``On certain incomplete statistics'' by Lima et al}

\author{M. Pezeril, \\ {\it Laboratoire de Physique de l'\'etat Condens\'e}, \\
{\it Universit\'e du Maine, 72000 Le Mans, France} \\
A. Le M\'ehaut\'e, and Q.A. Wang\\
{\it Institut Sup\'erieur des Mat\'eriaux et M\'ecaniques Avanc\'es}, \\
{\it 44, Avenue F.A. Bartholdi, 72000 Le Mans, France}}

\date{}

\maketitle

\begin{abstract}
Lima et al recently claim that ({\em Chaos, Solitons $\&$ Fractals,} 2004;19:1005)
the entropy for the incomplete statistics based on the normalization $\sum_ip_i^q=1$
should be $S=-\sum_ip_i^{2q-1}\ln_qp_i$ instead of $S=-\sum_ip_i^{q}\ln_qp_i$
initially proposed by Wang. We indicate here that this conclusion is a result of
erroneous use of temperature definition for the incomplete statistics.
\end{abstract}

{\small PACS : 05.20.-y, 05.70.-a, 02.50.-r}

\vspace{2cm}

In a recent work\cite{Lima03} addressing the incomplete statistics (IS) based upon
the normalization $\sum_ip_i^q=1$ proposed by Wang\cite{Wang01,Wang02c,Wang04a},
Lima et al calculated the entropy for IS with a method using the usual thermodynamic
relations preserved within the nonextensive statistical mechanics
(NSM)\cite{Tsal88,Curado,Tsal98} which is derived from Tsallis entropy
$S=-\sum_ip_i^{q}\ln_qp_i$ (let Boltzmann constant $k$ be equal to unity) where
$\ln_qp_i=\frac{p_i^{1-q}-1}{1-q}$ is the so called generalized logarithm, $q$ is
the nonadditive index and $p_i$ is the probability that the system of interest be
found at the state $i$. They ended with an entropy
$S_{IS}=-\sum_ip_i^{2q-1}\ln_qp_i$ and concluded that $S$ was intrinsically
connected to the complete distribution normalization $\sum_ip_i=1$.

Their calculation is based on the following relationships. 1) The first law :
$dU=TdS-PdV$ or $dF=-SdT-PdV$, $F=U-TS=-T\ln_qZ$, and $S=-\left(\frac{\partial F}{
\partial T}\right)_V$ where $T=1/\beta=\left(\frac{\partial U}{
\partial S}\right)_V$ is the temperature and $Z$ is the partition function
of IS. This method is correct as long as the above relationships hold, as indicated
by Lima et al. By this comment, we would like to indicate that these conventional
thermodynamic relationships are formally preserved within IS but with a deformed
entropy and a physical or generalized temperature $T_p$ different from the $T$ given
above. So the conclusion of \cite{Lima03} should be modified.

Due to the fact that there are several versions for NSM proposed from different
statistics or information considerations, thermodynamic functions do not in general
have the same nonadditive nature in different versions of NSM. This has led to
different definitions of physical temperature $\beta_p$ which is sometimes equal to
$\beta$\cite{Curado,Wang03c}, sometimes equal to $\beta$ multiplied by a function of
the partition function $Z^{q-1}$ or $Z^{1-q}$\cite{Wang02c}. Within IS using the
energy expectation $U=\sum_ip_i^qE_i$ in respecting its nonadditivity\cite{Wang02a},
the maximum of $S$ leads to the distribution function
$p_i=\frac{1}{Z}[1-(1-q)\beta_pE_i]^\frac{1}{1-q}$ with the partition function
$Z^q=\sum_{i}[1-(1-q)\beta_pE_i]^\frac{q}{1-q}$ where $\beta_p=1/T_p$ is given
by\cite{Wang02c} :
\begin{equation}                                        \label{3xx}
\beta_p=Z^{1-q}\frac{\partial{S}}{\partial{U}}=Z^{1-q}\beta.
\end{equation}
On the other hand, the introduction of the distribution $p_i$ into Tsallis entropy
gives
\begin{equation}                                        \label{7x}
S=\frac{Z^{q-1}-1}{q-1}+\beta_pZ^{q-1}U
\end{equation}
or $S_p=Z^{1-q}S=\ln_qZ+\beta_pU$ where $S_p$ is a ``deformed entropy'' introduced
in \cite{Wang02c} to write the heat as $dQ=T_pdS_p$ and the first law as
$dU=T_pdS_p+dW$ or $dF=-S_pdT_p+dW$ where
\begin{equation}                                        \label{8x}
F=U-T_pS_p=-T_p\ln_qZ
\end{equation}
is the Helmholtz free energy and $dW$ is the work done to the system. We get
\begin{equation}                                        \label{a9x}
S_p=-\left(\frac{\partial F}{\partial T_p}\right).
\end{equation}
Then, thanks to the techniques of \cite{Lima03}, this mathematically useful
``entropy'' whose probability dependence was unknown\cite{Wang02c} can be calculated
and given by :
\begin{equation}                                        \label{9x}
S_p=-\sum_ip_i^{2q-1}\ln_qp_i=-\sum_ip_i^{q}\frac{p_i^{q-1}-1}{q-1}.
\end{equation}
which is by definition not the original entropy $S$ of IS\cite{Wang01}. $S_p$ is
concave only for $q>1/2$ so that not to be maximized to get distribution functions
for NSM although its maximum formally leads to $p'_i\propto
[1-(q-1)\beta_pE_i]^\frac{1}{q-1}$. Notice that this latter is not the original
distribution function $p_i$ of IS. It should also be noticed that the method for
calculating $S_p$ cannot be used for calculating $S$ by using $\beta$ or $T$ within
this version of IS because $S\neq -\left(\frac{\partial F}{\partial T}\right)$
although we can formally write $F=U-T_pS_p=U-TS$. In addition, $Z$ is not derivable
with respect to $\beta$ since it is a self-referential function when written as a
function of $\beta$.

$S$ can be calculated in this way only when $\beta_p=\beta$, which is possible
within NSM only when ``unnormalized expectation'' is used\cite{Wang03c}. For IS, if
one use the unnormalized expectation $U=\sum_ip_iE_i$ (or normalized by
$U=\frac{\sum_ip_iE_i}{\sum_ip_i}$), the maximum of $S$ will lead to the
distribution function $p_i=\frac{1}{Z}[1-(q-1)\beta_pE_i]^\frac{1}{q-1}$ with the
partition function $Z^q=\sum_{i}[1-(q-1)\beta_pE_i]^\frac{q}{q-1}$. It is easy to
prove $\sum_{i}p_i=Z^{q-1}+(q-1)\beta_pU$ leading to
\begin{equation}                                        \label{a7x}
S=\frac{Z^{q-1}-1}{q-1}+\beta_pU
\end{equation}
which implies $\beta_p=\frac{\partial{S}}{\partial{U}}=\beta$. Now we can naturally
write $dQ=TdS$ and $dU=TdS+dW$ or $dF=-SdT+dW$ with $F=U-TS=-T\ln'_qZ$ where
$\ln'_qx=\frac{x^{q-1}-1}{q-1}$ is another generalized logarithm. It is easy to
prove that $S=-\left(\frac{\partial F}{\partial T}\right)$ with the normalization
$\sum_ip_i^q=1$ leads to Tsallis entropy.

Let us give here a summary of the definitions of temperature for different versions
of NSM found in the literature. We have $\beta_p=Z^{q-1}\beta=Z^{q-1}\frac{\partial
S}{\partial U}$ for the normalized expectations $U=\sum_ip_iE_i$ or
$U=\sum_ip_i^qE_i/\sum_ip_i^q$ with $\sum_ip_i=1$;
$\beta_p=Z^{1-q}\beta=Z^{1-q}\frac{\partial S}{\partial U}$ for the normalized
expectations $U=\sum_ip_i^qE_i$ with $\sum_ip_i^q=1$; and
$\beta_p=\beta=\frac{\partial S}{\partial U}$ if and only if unnormalized
expectation $U=\sum_ip_i^qE_i$ with $\sum_ip_i=1$ or $U=\sum_ip_iE_i$ with
$\sum_ip_i^q=1$ is used. The conclusion of this comment is that
$S=-\left(\frac{\partial F}{\partial T}\right)$ is valid only for $\beta_p=\beta$
and that Tsallis entropy can be derived with this formula for both $\sum_ip_i=1$ and
$\sum_ip_i^q=1$.

\section*{Acknowledgments}

We would like to thank J.A.S. Lima for useful discussions and for sending us
important references.

\end{document}